# Classification study of WISE infrared sources: identification of candidate asymptotic giant branch stars [*]

Xun Tu[1,2] and Zhong-Xiang Wang[1]

[1] Key Laboratory for Research in Galaxies and Cosmology and Shanghai Astronomical Observatory, Chinese Academy of Sciences, Shanghai 200030, China; *wangzx@shao.ac.cn*
[2] Graduate University of Chinese Academy of Sciences, Beijing 100049, China

**Abstract** In the Wide-field Infrared Survey Explorer (WISE) all-sky source catalog there are 76 million mid-infrared point sources that were detected in the first three WISE bands and have association with only one 2MASS near-IR source within $3''$. We search for their identifications in the SIMBAD database and find 3.2 million identified sources. Based on these known sources, we establish three criteria for selecting candidate asymptotic giant branch (AGB) stars in the Galaxy, which are three defined zones in a color-color diagram, Galactic latitude $|b| \leq 20°$, and "corrected" WISE third-band $W3_c \leq 11$. Applying these criteria to the WISE+2MASS sources, 1.37 million of them are selected. We analyze the WISE third-band W3 distribution of the selected sources, and further establish that $W3 \leq 8$ is required in order to exclude a large fraction of normal stars from them. We therefore find 0.47 million candidate AGB stars in our Galaxy from the WISE source catalog. Using $W3_c$, we estimate their distances and derive their Galactic distributions. The candidates are generally distributed around the Galactic center uniformly, with 68% (1-$\sigma$) of them within approximately 8 kpc. We discuss the idea that optical spectroscopy can be used to verify the C-rich AGB stars in our candidates, and thus a fraction of them (∼10%) will be good targets for the Large Sky Area Multi-Object Fiber Spectroscopic Telescope (LAMOST) survey that is planned to start in fall of 2012.

**Key words:** stars: AGB and post-AGB — infrared: stars — stars: carbon

## 1 INTRODUCTION

After being launched on 2009 December 14, the Wide-field Infrared Survey Explorer (WISE) has examined the entire sky at four mid-infared (MIR) bands, 3.4, 4.6, 12 and 22 μm (hereafter named W1, W2, W3 and W4, respectively), over a year (Wright et al. 2010). Depending on the number of coverages of a sky region by WISE, the 5-$\sigma$ point source sensitivities reached were approximately better than 0.08, 0.11, 1 and 6 mJy in the four bands, respectively. The FWHMs of the averaged point spread function for WISE in the four bands were $6.1''$, $6.4''$, $6.5''$ and $12.0''$. The WISE all-sky images and source catalog, released in 2012 March, contain over 563 million objects and provide a massive amount of information on MIR properties of many different types of celestial objects and their related phenomena (Wright et al. 2010). Given these, we have conducted analyses of the WISE sources for the purpose of identifying different classes of candidate objects among them. In this paper, we report our work on identification of candidate asymptotic giant branch (AGB) stars in the WISE catalog.

---

∗ Supported by the National Natural Science Foundation of China.

AGB stars are low- to intermediate-mass (1–8 $M_\odot$) stars that have reached their final evolutionary stage (e.g., Herwig 2005). They have intense mass loss via a super-wind, and emission from the ejected dust marks them as bright IR sources. Due to the so-called dredge-up process, O-rich or C-rich spectral features appear in emission from them, and these stars are called O-rich or C-rich AGB stars, respectively. These AGB stars, along with supernovae, are the major sources contributing to the enrichment of the interstellar medium (ISM) in galaxies (see, e.g., Matsuura et al. 2009). A survey of the Galactic AGB stars and studies of their properties will greatly help our understanding of chemical evolution of the ISM in the Galaxy. Previously, Ishihara et al. (2011) investigated the distributions of C-rich and O-rich AGB stars in our Galaxy using the AKARI MIR all-sky survey results (Murakami et al. 2007), but their results were limited by the relatively low sensitivities of the AKARI survey (5-$\sigma$ detection limits of 50 and 90 mJy at 9 and 18 μm respectively).

We describe our WISE data analyses in Section 2, which include WISE data selection, identification of WISE sources in the SIMBAD database, determination of the zone for AGB stars in an IR color-color diagram based on the AGB stars known from the SIMBAD database, selection of a spatial distribution criterion for excluding extragalactic sources, and calculation of "corrected" W3 magnitudes $W3_c$ for estimation of distances to and further selection of AGB stars in the Galaxy. Results and further analyses of the identified candidate AGB stars in the Galaxy are presented in Section 3. We discuss our results in Section 4.

## 2 DATA ANALYSIS

### 2.1 WISE Data

The released WISE all-sky source catalog was downloaded from the NASA/IPAC infrared science archive (IRSA). The catalog mainly contains positions, photometry and uncertainties in the four WISE bands for the sources detected by WISE. The positions were calibrated to the 2MASS point source catalog, thus generally having an accuracy of $\sim 0.2''$ on each axis. Both profile-fitting and aperture photometry were performed to measure brightnesses of the WISE sources; we used the measurements from profile-fitting photometry in this work. The detailed description of the data analyses and products can be found in the WISE All-Sky Release Explanatory Supplement[1].

Very approximate identification of the WISE sources can be conducted by directly constructing a color-color diagram (Wright et al. 2010). In order to as clearly as possible identify different classes of objects, we followed the method used by Ishihara et al. (2011). The WISE catalog contains information about whether sources are associated with the 2MASS All-Sky point source catalog (Skrutskie et al. 2006), in which a $3''$ radius was used for determining the association. In order to be able to define classes of objects with sufficient color information, we selected those WISE sources that are associated with one 2MASS source within $3''$ and were detected by all W1, W2 and W3 bands. We noted that to avoid ambiguous identification, 1.02 million WISE sources associated with more than one 2MASS source were excluded from the selection. Because in general these 1.02 million sources have Galactic latitudes $|b|<20°$, they are likely to be stars. Additionally excluding 6.23 million sources that are marked in any one of the WISE bands with a cc_flag denoting severe artifact contamination or confusion, we found that there are 76.10 million WISE sources that have both association with one 2MASS point source and the W1–W3 three band flux measurements.

---

[1] *http://wise2.ipac.caltech.edu/docs/release/allsky/expsup*

### 2.2 SIMBAD Identifications

In order to identify different classes of sources and thus select candidate AGB stars from their color-color and/or color-magnitude information, we searched for known sources among 76 million WISE sources. We input the positions of the 76 million sources in the SIMBAD database with a search radius of $10''$, and found 3.4 million sources with identifications. Among them, 2.5 million cases are stars and over 40% of the remaining ones are galaxies. We carefully examined the distributions of position differences for these matches, since our search radius was $10''$ and this radius could cause false identifications, particularly in the Galactic plane. We found that for most of the classes, distribution peaks below $1''$ are clearly visible, likely indicating true matches between the WISE and SIMBAD sources. For a few classes of sources, their distributions have a rising tail at distances greater than $2''$ and this tail is likely caused by positional coincidence when the density of sources is sufficiently high. We therefore rejected identifications with greater than $2''$ in radius for every class of objects.

In Table 1, the classes that have more than 1% SIMBAD-identified sources are listed for illustration. The numbers of different classes resulting from the criterion of the separation distances being less than $2''$ are given in column 3 of Table 1. The total number of them is 3.2 million.

**Table 1** Numbers of IR Sources Identified in the SIMBAD Database

| Class | Number ($r \leq 10''$) | Number ($r \leq 2''$) | Percent ($r \leq 2''$) |
|---|---|---|---|
| Star | 2512531 | 2455250 | — |
| Total | 886300 | 748815 | 100 |
| Galaxy | 348533 | 325390 | 43 |
| IR | 78617 | 28305 | 3.8 |
| $*$inCl | 63327 | 57473 | 7.7 |
| PM$*$ | 56281 | 54441 | 7.3 |
| GinCl | 34787 | 32568 | 4.3 |
| $*$in$**$ | 29024 | 26199 | 3.5 |
| V$*$ | 28764 | 25167 | 3.4 |
| LPV$*$ | 26188 | 24661 | 3.3 |
| Radio | 23930 | 10509 | 1.4 |
| semi-regV$*$ | 14819 | 14539 | 1.9 |
| QSO | 14640 | 13311 | 1.8 |
| GinGroup | 13297 | 12927 | 1.7 |
| C$*$ | 12516 | 11039 | 1.5 |
| Mira | 8422 | 8227 | 1.1 |
| low-mass$*$ | 8420 | 7896 | 1.1 |
| $**$ | 8555 | 7820 | 1.0 |

Notes: a) Stars are counted separately because of their dominantly large number.
b) Sign '$*$' denotes star; for example $**$ denotes the class of double or multiple star.

### 2.3 Color-Color Classification

Using the 3.2 million known sources, we tested different ways of separating the AGB stars from the other sources (see, e.g., van der Veen & Habing 1988; Blum et al. 2006; Lagadec et al. 2007; Ishihara et al. 2011), and found that the color-color diagram of $J - Ks$ versus $Ks-$W3 generally worked. In Figure 1 we show the diagram for the known AGB stars as well as contours that define the regions for the stars and galaxies. As can be seen, the AGB stars are generally separated from the other sources, although a fraction of them are located in the region of the stars and slightly overlap with the blue end of the galaxy region. Because of the high density of the stars (Table 1), we defined the following three zones (zone 1, 2 and 3, respectively) that include most of the AGB stars while excluding the 2-$\sigma$ contour regions of the stars and galaxies:

$$0.42 + 0.8 \times (Ks - W3) < (J - Ks) < 0.72 + 1.66 \times (Ks - W3)$$
$$\text{for} \quad 0.35 < (Ks - W3) < 1.45, \quad (1)$$

$$0.71 + 0.6 \times (Ks - W3) < (J - Ks) < 2.25 + 0.6 \times (Ks - W3) \quad (2)$$

and

$$1.29 + 0.2 \times (Ks - W3) < (J - Ks) < 0.71 + 0.6 \times (Ks - W3)$$
$$\text{for} \quad (Ks - W3) > 1.45. \quad (3)$$

We checked contamination by other classes of the sources, i.e., how many other sources are located in the three zones. The results are summarized in the second column of Table 2; only sources with relatively large numbers ($> 500$) are listed. It can be noted that because of large numbers of a few classes of objects (particularly the stars and galaxies), even a small fraction of them can be dominant over the targets. For example, there are approximately 4000 galaxies in the three zones. We therefore added one selection criterion based on source positions, requiring $|b| \leq 20°$. This constraint, estimated from the $b$ distributions of the sources, helped exclude most of the extragalactic sources including a large number of variable stars in the Large and Small Magellanic Clouds (LMC and SMC; see Fig. 2). In addition, we noted that a large fraction of nearby stars, for example among the classes of the stars and proper motion stars, were also excluded (see Table 1). Adding this constraint, we re-calculated the numbers of the sources in the three zones and summarized the results in column 3 of Table 2.

### 2.4 WISE Color-Magnitude Analysis

The observed MIR fluxes and colors of an AGB star are generally determined by the interstellar extinction, mass loss rate of the star and distance. Since the extinction is usually negligible at the MIR wavelengths (Indebetouw et al. 2005), the colors are a function of only the mass loss rate. A relation between values of one IR color and mass loss rates is often found when the mass loss rates can be obtained (e.g., Matsuura et al. 2009; Zhang et al. 2010; Ishihara et al. 2011). Also if considering that luminosities of AGB stars are similar, distances to them can be estimated (e.g., Le Bertre et al. 2001). We therefore checked the WISE color-magnitude diagram of the SIMBAD sample of the AGB stars (bottom left panel in Fig. 3), in which we also included the known AGB stars in the LMC and SMC.

We found that the C-rich stars (few O-rich stars in the SMC were found in our sample; see, e.g., Marshall et al. 2004) in the SMC can generally be represented by a line in the diagram. Using a "robust" least absolute deviation method, we fit W1−W3 and W3 to a linear model and found W3=11.53−1.83(W1−W3). This dependence on W1−W3 presumably reflects the mass loss rates of the stars and can probably be removed by defining a "corrected" magnitude $W3_c$=W3+1.83(W1−W3). In the top left panel of Figure 3, we show the W1−W3 and $W3_c$ color-magnitude diagram of the AGB stars in the Galaxy and SMC, and as can be seen, the distribution of most SMC stars is now corrected to be horizontally elongated, with the central $W3_c$ value likely indicating the distance to the SMC and the spread of $W3_c$ values the depth of the SMC and probably the differences in the mass loss rates as well. We therefore used a distance of 61 kpc to the SMC (Lagadec et al. 2007) and derived the corresponding distance based on $W3_c$, and found most of the AGB stars in the Galaxy were in a range of 4–16 kpc (Fig. 3). There are a few outliers in our sample of the Galactic AGB stars that have $W3_c > 15$, which are probably the AGB stars found in nearby dwarf galaxies by the previous surveys. As a check of our distance estimate method, we searched the

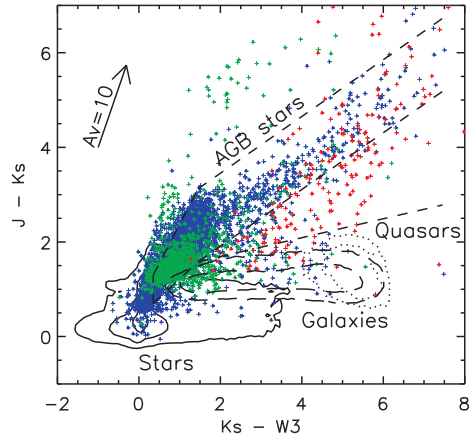

**Fig. 1** $J - Ks$ versus $Ks-W3$ color-color diagram of the known C-rich and O-rich stars (*blue*: C-rich AGB stars; *green*: AGB stars; *red*: O-rich stars). The solid contours show 1-, 2-, and 3-$\sigma$ regions of the normal stars, long-dashed contours 1- and 2-$\sigma$ regions of the galaxies, and dotted contours 1- and 2-$\sigma$ regions of the quasars. The dashed lines define the region for the AGB stars, although they are slightly contaminated by other sources. The direction of extinction is shown by the arrow.

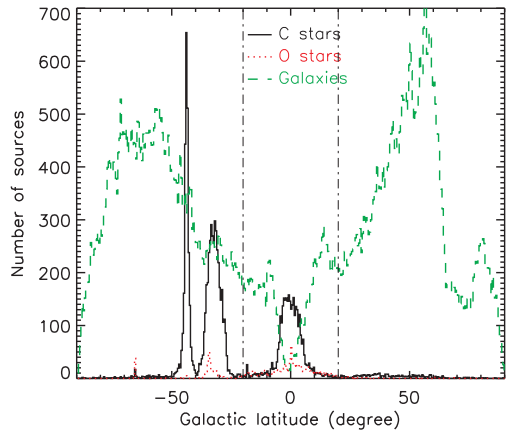

**Fig. 2** A comparison of $b$ distributions of the AGB stars and that of the galaxies. The latter is scaled by 3 to clearly show the comparison. Large numbers of the known C-rich AGB stars in the LMC and SMC are clearly visible.

C-rich and O-rich stars reported in Le Bertre et al. (2003) from the WISE catalog and found 108 and 113 of them, respectively. The distances to these stars were estimated by Le Bertre et al. (2003), with a factor of two in uncertainty. Their values were compared with that derived from our method, and 68% ($1\sigma$) of the 221 stars are consistent within 40%. The comparison result indicates that $W3_c$ can be used to estimate distances to AGB stars, although we cautiously note that the "systematic" uncertainty of the distances given by Le Bertre et al. (2003) is large and we have assumed that the intrinsic properties of the AGB stars through the WISE bands in the Galaxy and SMC are very similar.

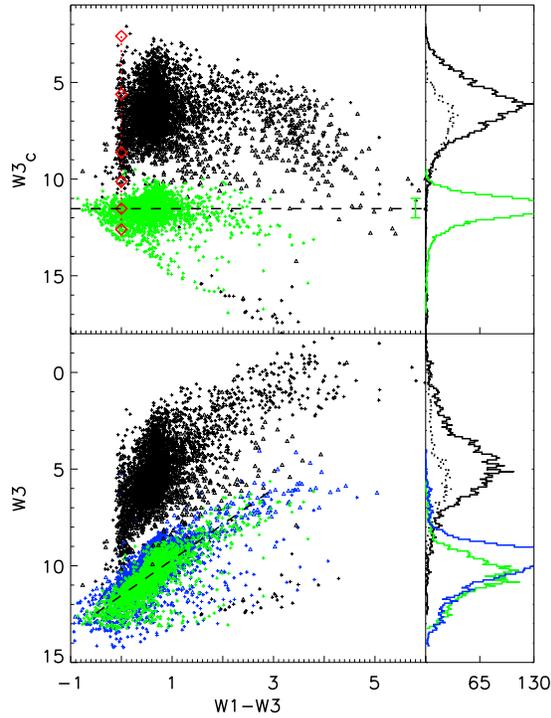

**Fig. 3** *Bottom* panels: W3 versus W1−W3 diagram of the C-rich and O-rich AGB stars in the Galaxy (black plus and triangle symbols, respectively), LMC (blue plus and triangle symbols), and SMC (green plus signs). The C-rich stars in the SMC can generally be represented by the dashed line. The distributions of these stars as a function of W3 are respectively shown in the right narrow panel. *Top* panels: $W3_c$ versus $W1-W3_c$ diagram of the same stars in the Galaxy and SMC. The red diamonds indicate distances of 1, 4, 16, 32, 61 and 100 kpc, estimated from $W3_c$ of the C-rich stars in the SMC (dashed line). The distributions of the stars are also shown in the right panel, which are sharpened compared to those in the bottom right panel. The green bar marks the standard deviation of 0.5 mag from $W3_c = 11.5$ of the AGB stars in the SMC.

From the above analysis, an additional criterion for selecting AGB stars in the Galaxy can be set. As shown in Figure 3, more than 80% of the AGB stars in the Galaxy have $W3_c \leq 11$ (or distance less than 50 kpc). Thus by requiring $W3_c \leq 11$, we further significantly reduced contamination from galaxies and excluded stars that are not located in our Galaxy. The results are summarized in columns 4–7 of Table 2.

Following Ishihara et al. (2011), we divided the classes listed in Table 2 into three groups: A) objects that are classified as different categories, but can be considered as the same group of our targets; B) objects are classified as other categories; C) classified targeted objects. We estimated the reliability of our classification using $(n_a + n_c)/(n_a + n_b + n_c)$. In group A, IR and variable stars were included as $n_a$ for both C-rich and O-rich stars, while Mira and candidate AGB stars were included only as O-rich stars. The AGB and OH/IR stars were combined together as O-rich stars (Ishihara et al. 2011). The results are given in Table 2. The reliabilities for the C-rich and O-rich stars satisfying all the selection criteria are 68% and 71%, respectively.

Table 2 Numbers of Sources in the Three Defined Zones of the AGB Stars

| Class | Total (All-sky) | Total ($|b|\leq 20°$) | Zone 1 | Zone 2 | Zone 3 | Total |
|---|---|---|---|---|---|---|
| Group A (same) | | | | | | |
| IR | 13194 | 12521 | 3022 | 2800 | 1481 | 7303 |
| V* | 8718 | 8408 | 4509 | 2044 | 1084 | 7637 |
| Semi-regV* | 8599 | 2409 | 1903 | 220 | 137 | 2260 |
| Mira | 3570 | 2244 | 1120 | 510 | 482 | 2112 |
| Candidate_AGB* | 2075 | 2070 | 0 | 803 | 304 | 1107 |
| PulsV* | 1467 | 1213 | 1084 | 39 | 26 | 1149 |
| Group B (other) | | | | | | |
| Star | 9151 | 4976 | 3647 | 353 | 89 | 4089 |
| Candidate_YSO | 1759 | 1742 | 58 | 87 | 72 | 217 |
| PM* | 9351 | 1690 | 1021 | 1 | 0 | 1022 |
| *inCl | 1347 | 1033 | 553 | 24 | 7 | 584 |
| Galaxy | 3870 | 903 | 20 | 1 | 7 | 28 |
| YSO | 745 | 709 | 66 | 108 | 61 | 235 |
| Group C (targeted) | | | | | | |
| C* | 8611 | 3259 | 2053 | 698 | 51 | 2802 |
| AGB* | 797 | 615 | 498 | 84 | 27 | 609 |
| OH/IR | 192 | 189 | 1 | 62 | 77 | 140 |
| Fraction (%) | | | | | | |
| C* | 55 | 63 | 64 | 74 | 71 | 68 |
| AGB*+OH/IR | 53 | 68 | 62 | 84 | 93 | 71 |

Columns (4)–(7) are the numbers of sources that additionally satisfy $W3_c \leq 11$.

## 2.5 WISE Color-Color Separation Between C-Rich and O-Rich Stars

We investigated different combinations of WISE band color-color analyses for the purpose of separating C-rich and O-rich stars, and did not find a method that can clearly discriminate between the two sets of the stars. However as shown in Figure 4, which is the WISE $W2-W3$ versus $W1-W2$ diagram of the known C-rich and O-rich stars in our sample, the stars with extremely red colors can generally be defined. The conditions $W2-W3>1$ for $W1-W2<1$ and $W2-W3\geq 0.75(W1-W2)+1$ for $W1-W2\geq 1$ can separate the OH/IR stars and a small fraction of the AGB stars from the C-rich stars in our sample.

## 3 RESULTS

Applying the IR color-color, Galactic latitude and $W3_c \leq 11$ selection criteria to the 76 million WISE+2MASS sources, we searched for candidate AGB stars and found 1.37 million sources satisfying the three criteria, with 1.16, 0.189, and 0.025 million sources in zones 1, 2, and 3, respectively. The distributions of the selected sources in the three zones as a function of W3 are shown in Figure 5. In addition, the same distribution for all 76 million sources is also shown in the figure. As can be seen, different from the distribution of the 76 million sources, there appear to be two components in both the zone 1 and 2 distributions. We noted that for the zone 1 sources, the upper edge of the main peak (at W3=10) was caused by the requirement of $W3_c \leq 11$. On the basis of the known AGB stars in the SMC, $W3_c \approx 11.53$ or $W3 \approx 11.53 - 1.83(W1-W3)$. Most AGB stars in the Galaxy and SMC have $W1-W3$ colors of 0–1 mag (Fig. 3), implying a range of 5.3–7.2 mag for a distance of 8 kpc or 3.8–8.6 mag for a distance of 4–16 kpc. Therefore the components of faint sources in the zone 1 and 2 distributions ($W3\geq 8$) are likely to be normal stars, not AGB stars.

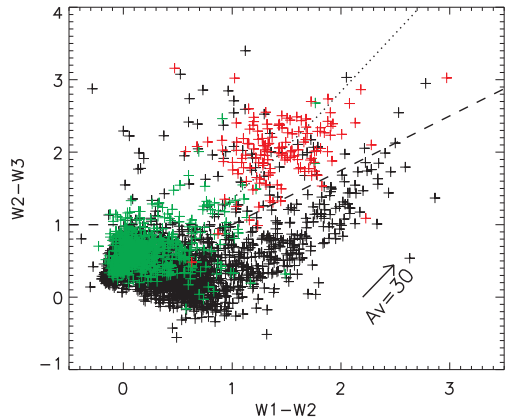

**Fig. 4** WISE color-color diagram of the known C-rich stars (*black*), AGB stars (*green*), and OH/IR stars (*red*). The dotted curve indicates the colors for objects at different low temperatures (Wright et al. 2010), indicating that the OH/IR stars have temperatures of 500–800 K. The dashed lines can generally separate the OH/IR and a small fraction of the AGB stars from the C-rich stars in our sample.

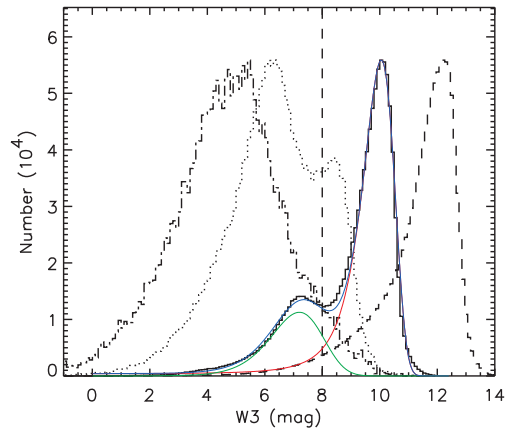

**Fig. 5** Distributions of the candidate AGB stars in the zones 1 (*solid curve*), 2 (*dotted curve*), and 3 (*dash-dotted curve*) and 76 million WISE sources (*dashed curve*) as a function of W3. The distributions of the latter three cases were respectively scaled to best show the comparisons among them. Two components are visible in the zone 1 and 2 curves with a boundary approximately at 8 mag. The two components can be well represented by a modified Gaussian function (green curve for the AGB component, red for the normal star component, and blue for the sum of the two in zone 1).

Using W3=8 as an approximate boundary to separate the candidate AGB stars from normal stars in the zone 1 and 2 sources, 0.295 and 0.151 million candidate AGB stars were obtained. Now including those in zone 3, the total number is 0.471 million (34% of the 1.37 million sources). We also tested fitting the two components in the distribution curves with a modified Gaussian function $y \sim \exp(-x^2/2\sigma^2)$, where $x$ was set as a function of $\exp(W3)$, and the fitting results of the numbers of candidate AGB stars and normal stars were consistent with those from simply using W3=8 as a separation boundary. It should be noted that while the fractions of two components can be more accurately estimated by fitting, the sources cannot be accordingly separated.

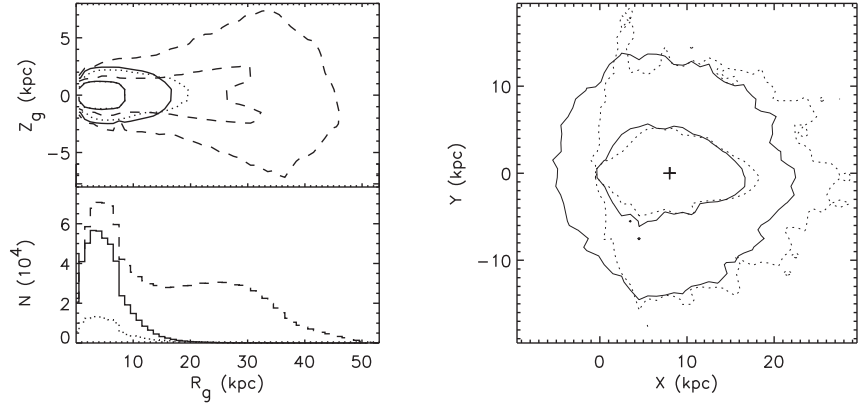

**Fig. 6** Galactic distributions of the candidate AGB stars. *Top left:* 1- and 2-$\sigma$ contour regions (*dashed curves*) of the 1.37 million candidates. The solid and dotted contours show the same contours for the further selected 0.47 million candidates (W3≤8) and for those candidate O-rich stars obtained from the WISE color-color analysis, respectively. *Bottom left:* Distributions of the total candidates (*dashed*), further selected candidates (*solid*), and those candidate O-rich stars (*dotted*) along the Galactocentric distance. *Right:* 1- and 2-$\sigma$ distributions of the further selected candidate AGB stars (*solid curves*) and O-rich stars (*dotted curves*). The Sun is located at $X = 0$, $Y = 0$, and the Galactic center is marked by the plus sign.

In Table 2 we estimated that approximately 20% of the selected sources should belong to Group B, most of which should be normal stars. Our study of the W3 distributions has shown that the fraction of normal stars is as high as 66%. This large difference was likely caused by the star sample in the SIMBAD database, whose distribution along W3 peaks at 10 mag. Comparing it to that of the 76 million sources (peaks at 12 mag, which corresponds to 0.5 mJy; Fig. 5), a large fraction of stars that have W3=10–12 were not represented by the SIMBAD sample. This bias probably made the fraction of normal stars from our selection method significantly underestimated.

Galactic reddening towards each selected source was estimated using the dust map of the Galaxy (Schlegel et al. 1998) and the reddening laws for the WISE bands given from Weingartner & Draine (2001) and Li & Draine (2001) were used. The distances were estimated using dereddened W3$_c$ and their Galactocentric distances $R_g$ and vertical distances from the Galactic plane $Z_g$ were accordingly estimated as well (a distance of 8 kpc to the Galactic center was used; Gillessen et al. 2009). The distributions of the sources along $R_g$ and in the $R_g$ versus $Z_g$ plane are shown in the left panel of Figure 6. As can be seen, the 2-$\sigma$ contour of the distribution in the latter extends to $R_g \approx 45$ kpc and $Z_g \approx 7$ kpc, both unrealistically large, indicating there is a large fraction of normal stars in our selected candidate AGB stars.

The distances of normal stars were largely over-estimated using W3$_c$. Reducing our total number of candidates by 66% by requiring W3≤8, we re-plotted the distribution contours in Figure 6. The maximum $R_g$ and $Z_g$ values were changed approximately to 17 and 1 kpc, respectively. These values were significantly reduced and comparable to the typical size of the Galaxy. In the right panel of the figure, the distribution of the further selected candidate AGB stars projected onto the Galactic plane is shown. The candidates are generally distributed around the Galactic center.

Using the WISE color-color analysis given in Section 2.5, we searched for candidate O-rich stars among the 0.471 million candidates and found 0.119 million. The distributions of these 0.119 million sources are shown in Figure 6, and no significant differences between them and those of the other candidates are seen. This suggests that we probably did not select O-rich stars successfully, given that the currently known O-rich AGB stars appear to concentrate towards the Galactic center (Ishihara et al. 2011). The result is expected because as discussed in Ishihara et al. (2011), only the AKARI 9 and 18 µm bands well covered the silicate features of O-rich stars and clearly distinguished between O-rich and C-rich stars.

## 4 DISCUSSION

We analyzed the data in the newly released WISE all-sky source catalog to search for C-rich and O-rich AGB stars in the Galaxy. We selected 76 million point sources in the catalog that were detected in the first three WISE bands and have only one 2MASS counterpart within $3''$.

A color-color diagram for these WISE+2MASS sources was constructed and zones in the diagram for AGB stars were defined. The spatial requirement of $|b| \leq 20°$ was established to help exclude extra-galactic sources. Based on the known AGB stars in the SMC, we also derived "corrected" WISE third-band magnitudes $W3_c$ to estimate distances for AGB stars and to select candidates.

Applying the color-color, spatial, and $W3_c$ criteria, 1.37 million candidates were selected. Analyses of them have shown that $W3 \leq 8$ was further required to separate the AGB stars from the normal stars in the candidates. As a result, 0.47 million candidate AGB stars were found, with their distances estimated and distributions in the Galaxy obtained. Given the sensitivities of the WISE survey, which are an order of magnitude deeper than the MIR brightnesses expected for AGB stars in the Galaxy, these candidates probably constitute a large fraction of them. Observations of them are warranted for verifications.

In order to determine the AGB nature of the candidates, MIR spectroscopy best serves the purpose since the continuum of such a spectrum contains information about the mass loss rate of a target and the spectral features help directly identify whether it is a C-rich or O-rich star (Kraemer et al. 2002; Sloan et al. 2003). Currently only the Stratospheric Observatory for Infrared Astronomy (SOFIA) is operating at MIR wavelengths with spectroscopic capability, but a few large ground-based telescopes can reach a sensitivity of $\sim$10 mJy at the wavelengths. However, neither is suitable for a spectroscopic survey of 0.5 million sources.

We note that strong optical spectral features arise from carbon-rich molecules (such as CN, $C_2$, and CH; Barnbaum et al. 1996) and C-rich stars can be identified from optical spectroscopy. The AGB stars in the SIMBAD database have a $V - K$ distribution with the peak at $V - K \approx 7$ and the $Ks$ distribution of our 0.47 million candidates peaks at $Ks \sim 8$. Most of our candidates thus could have $V \sim 15$. Including reddening, a fraction ($\sim$10%) of the candidates are good targets for the Large Sky Area Multi-Object Fiber Spectroscopic Telescope (LAMOST, also called the Guo Shou Jing Telescope) in China, which will start operation from the fall of 2012.

**Acknowledgements** This work was supported in part by the Pre-phase Studies of Space Science Projects of the Chinese Academy of Sciences, National Basic Research Program of China (973 Project, 2009CB824800), and the National Natural Science Foundation of China (Grant No. 11073042). ZW is a research fellow of the One-Hundred-Talents project of Chinese Academy of Sciences. This publication makes use of data products from the Wide-field Infrared Survey Explorer, which is a joint project of the University of California, Los Angeles, and the Jet Propulsion Laboratory/California Institute of Technology, funded by the National Aeronautics and Space Administration.